# Observation of intrinsic crystal phase in bare CrI$_3$ ferromagnetism

*Zhen Liu[1,2], Yongzheng Guo[1,2], Zhiyong Chen[1,2], Tao Gong[1,2], Yue Li[1,2], Yuting Niu[1,2], Yingchun Cheng[3], Haipeng Lu[1,2], Longjiang Deng[1,2,\*] and Bo Peng[1,2,\*]*

[1]National Engineering Research Center of Electromagnetic Radiation Control Materials, School of Electronic Science and Engineering, University of Electronic Science and Technology of China, Chengdu 611731, China

[2]State Key Laboratory of Electronic Thin Films and Integrated Devices, University of Electronic Science and Technology of China, Chengdu, 611731, China

[3] Key Laboratory of Flexible Electronics & Institute of Advanced Materials, Jiangsu National Synergetic Innovation Center for Advanced Materials, Nanjing Tech University, Nanjing 211816, China

\*Correspondence author's e-mail: denglj@uestc.edu.cn; bo_peng@uestc.edu.cn

**Abstract:** Intrinsic structural phase is a crucial foundation for the fundamental physical properties, and for creating innovative devices with unprecedented performances and unique functionalities. Long-range ferromagnetic orders of van der Waals $CrI_3$ are strongly tied with interlayer stacking orders. However, the intrinsic structure of few-layer $CrI_3$ still remains elusive; the predicted monoclinic phase has not yet been experimentally detected in bare few-layer $CrI_3$. Here we uncover the intrinsic structure of few-layer $CrI_3$ with interlayer antiferromagnetic coupling, which unambiguously show monoclinic stacking in both bare and hBN-encapsulated bilayer and tri-five-layer $CrI_3$ throughout an entire temperature range from 300 to 10 K. An exotic spring damping effect from hBN encapsulation layers is experimentally observed in hBN/$CrI_3$/hBN heterostructures, which partly hinders interlayer sliding of $CrI_3$. This work demonstrates the intrinsic monoclinic crystal phase of few-layer $CrI_3$ and associated correlation with magnetic orders, opening up numerous opportunities for creating magnetic texture by stacking design.

# 1 Introduction

Lattice structures are ubiquitous in nature, which determine diverse physical and chemical properties of materials. Exploring and controlling crystal structures is a central task of material engineering.[1-4] Lattice phase transition is considered as a significant approach to manipulate and control functionalities, and thus, understanding the underlying mechanism of phase transition is a basic premise and guarantee for technological applications.[5, 6] In particular, there are ample lattice phases existing in two-dimensional (2D) materials, which enable unique properties in electrical, optical, magnetic and catalysis.[7, 8] Structural phase transition can be achieved in many ways, such as chemical,[9] thermal,[10] strain,[11] laser heating,[12] electro-static doping[13] and electric field,[14] *etc*.

The emergence of 2D ferromagnet has opened up new horizons for engineering structural phase transition with magnetic orders together beyond the reach of existing materials.[15-17] The magnetic ground-state is coupled with structural stacking orders originating from interlayer orbital hybridization.[18, 19] Intricate phase transition behaviors-crystal phase transition and magnetic phase transition, have been theoretically predicted and experimentally observed in 2D ferromagnets. Control of magnetism by tuning interlayer stacking order has been realized through the hydrostatic pressure method. The lattice phase converts from monoclinic to rhombohedral stacking in hBN-encapsulated few-layer $CrI_3$ under high pressure, accompanied with a transition from interlayer anti-ferromagnetic (AFM) to ferromagnetic (FM) coupling.[20, 21] In addition, it is experimentally observed that the change of localized lattice structure will affect the corresponding magnetic order in hBN-encapsulated few-layer $CrI_3$, but the corresponding lattice phase is unclear.[22]

Typical 2D ferromagnetic $CrI_3$ bulk undergoes a crystal phase transition process from a high-temperature monoclinic ($C2/m$) phase to a low-temperature rhombohedral ($R\bar{3}$) phase at a critical temperature of ∼210 K, and long-range FM ordering is persisted up to ~61 K.[23, 24] Monoclinic and rhombohedral crystal phases show the same intralayer atomic arrangement but different interlayer stacking order. Each layer is laterally shifted by a translation vector of (1/3, -1/3) and (1/3, 0) with respect to the neighboring layer for rhombohedral and monoclinic stacking, respectively.[20, 25] However, theory has predicted that low-temperature crystal structure of few-layer $CrI_3$ is monoclinic, which differs from bulk $CrI_3$.[18, 26-31] It should be emphasized that, to date, all reported monoclinic $CrI_3$ few layers in the experiment are encapsulated by hBN protection layers.[20, 21, 32-34] The details of the reported works on structural identification of $CrI_3$ are summarized in Table 1 of supplementary information. A hypothesis has been proposed that few-layer $CrI_3$ is kinetically trapped in the room-temperature monoclinic phase by hBN encapsulation layers upon cooling, which block the structural phase transition.[20] However, this hypothesis is unproven and remains controversial, because hBN is only used as a protective layer in traditional cognition, and the influence on structural phase transition is unconsidered. Therefore, the intrinsic structure of bare $CrI_3$ few-layer still lacks direct experimental evidences. Moreover, the observation of intrinsic structure of bare few-layer $CrI_3$ is a pivotal challenge in the experiments and has not been achieved yet.

Here, we evidently uncover the intrinsic structure of $CrI_3$ through systematically studying bare, half-bare-half-encapsulated and all-hBN-encapsulated bilayer (2L) and tri-five-layer (3-5L) $CrI_3$; all of them show monoclinic stacking throughout a whole temperature range from 300 to 10 K. The experimental results validate that monoclinic stacking of few-layer $CrI_3$ is inherent, rather than resulting from hBN encapsulation, which overturn the present hypothesis and define the intrinsic structures of $CrI_3$. Alternatively, hBN encapsulation layers are validated to lead to a spring damping effect and partly diminish interlayer sliding for preserving monoclinic stacking, breaking out the conventional wisdom on the role of hBN encapsulation layers in 2D heterostructures.

## 2 Results and discussion

### 2.1 Spring damping effect of hBN on structural phase transition

In Ising ferromagnet $CrI_3$, the Cr atoms in each layer are bonded with six coordination I atoms to form an octahedral structure and the adjacent Cr atoms are in an arrangement of honeycomb structure (Figure 1(a)). Due to the strong magneto-crystalline anisotropy, long-range intrinsic ferromagnetism can be stabilized down to the limit of a single layer. Notably, the easy magnetization axis is out-of-plane for single-layer and anti-ferromagnetically aligned in adjacent layers, which gives rise to layer-dependence behaviors of few-layer $CrI_3$.[15, 35] Figure 1(b) shows the crystal structure schematics of hBN-encapsulated and bare few-layer $CrI_3$. The hBN encapsulation layers lead to spring damping to hold $CrI_3$ in monoclinic phase, however, this hypothesis has not yet been proven experimentally. A set of contrast experiments on the same 3L $CrI_3$ is investigated through angle-resolved polarized Raman spectroscopy, which is considered as a powerful tool to distinguish crystal structures.[20, 36-38] A half of the 3L $CrI_3$ flake is encapsulated by hBN layers and another half is bare. Figure 1(c) and supplementary information Figure S2 show that the Raman mode energy differences between the parallel (XX) and crossed (XY) configurations on bare and hBN-encapsulated $CrI_3$ exhibit different temperature dependence behaviors. The extracted slope of bare $CrI_3$ (~9.5x$10^{-4}$ $cm^{-1}$/K) is more than twice larger than that of the hBN-encapsulated $CrI_3$ (~4.4x$10^{-4}$ $cm^{-1}$/K), implying that the hBN layers introduce a spring damping effect and partly mitigate interlayer sliding for preserving monoclinic stacking throughout the whole temperature. This result breaks out the conventional understandings that hBN encapsulation layers do not influence the structural phase of encapsulated components in 2D heterostructures, and validates that hBN encapsulation layers play a significant role for the structural phase of heterostructures, particularly for phase change 2D materials.

Comprehensive studying of the intrinsic crystal structure and magnetic phase and associated correlation is the foundation for understanding fundamental physics of bare $CrI_3$ ferromagnetism. First of all, we focus on studying the intrinsic crystal structure of bare few-layer $CrI_3$, and confirming whether the crystal structure of few-layer $CrI_3$ remains unchanged with temperature or undergoes crystal phase transitions like bulk $CrI_3$. The influence of hBN on the crystal transition process will be discussed below. In the rhombohedral phase, according to the space group symmetry theory, the $E_g$ mode of 107 $cm^{-1}$ is doubly degenerated and the energy is independent of polarization angle.

As a result, the $E_g$ Raman features at XX and XY polarization channels coincide completely. But the symmetry of monoclinic phase is lowered, the $E_g$ mode splits into $A_g$ and $B_g$ modes with distinct selection rules, which can only be observed in the XX and XY polarization channels, respectively. Meanwhile, the polarization angle dependence is switched to a fourfold dependence.[20, 36] Therefore, the lattice structure can be determined by angle-resolved polarized Raman spectra. Figure S1(a) shows the room-temperature Raman spectra of bare 4L $CrI_3$ collected in the XX and XY channels. Five Raman active $A_g$ modes in the XX channels and three Raman active $B_g$ modes in the XY channels are observed. The monoclinic phase is confirmed by an emergence of splitting $A_g$ and $B_g$ modes near ∼107 cm$^{-1}$. Besides, room-temperature non-polarized Raman spectra of bare 3-5L $CrI_3$ are exhibited in Figure S1(b). There are five modes in each spectrum and the peak intensity ratio is almost the same.

## 2.2 Intrinsic monoclinic phase of bare bilayer $CrI_3$

Bilayer $CrI_3$ is the smallest basic unit with interlayer stacking order and an unambiguous understanding of its crystal structure is vital to reveal the essence of intrinsic structure[18, 32, 39]. For few-layer $CrI_3$, saturation magnetization can be achieved by applying an out-of-plane magnetic field of 2 T. The layer-number and spin-flip states can be identified by reflectance magneto-circular dichroism (RMCD) spectroscopy.[40] Figure 2(a) shows the out-of-plane magnetization of 2L $CrI_3$ taken at 10 K. A zero net magnetization can be observed within an external magnetic field of 1.2 T, corresponding to the interlayer AFM-coupled state. As the magnetic field is further increased across 1.2 T, a spin-flip takes place, corresponding to a meta-magnetic transition from layered AFM state to out-of-plane co-parallel FM spin state. In order to assure the accuracy of layer-number identification, a retested RMCD spectrum with a magnetic field scope of ±1 T has been done, shown in the inset of Figure 2(a). There is no detectable hysteresis loop or spin-flip state, further indicating that the $CrI_3$ flake is bilayer. It should be noted that the coercive force field here is larger than previously reported literatures, which is possibly related to the grain boundary distribution, density, orientation and interlayer interaction of the samples,[41, 42] reflecting the differences of the samples. Next, we utilize the polarized Raman spectroscopy to analyze the intrinsic structure of bare 2L CrI$_3$. At 295 K, the Raman mode near 105 cm$^{-1}$ shows a fourfold dependence on polarization angle (Figure 2(b)), which validates that bare 2L CrI$_3$ is monoclinic phase at room temperature, which is consistent with bulk. The stacking order is closely related to the magnetism of few-layer CrI$_3$, thus, it is necessary to study the crystal structure in non-magnetized and magnetized states. Temperature-dependence RMCD spectra are detected to confirm the magnetization state (Figure S3). When the temperature is cooled down to ~50 K, the bare 2L CrI$_3$ changes from a paramagnetic (PM) state to an AFM state, indicating a Néel temperature of 45 K in 2L CrI$_3$. Further considering the reported crystal transition temperature of ~210 K in bulk, we choose a temperature of 90 K as an exemplification. As shown in Figure 2(c), a clear fourfold dependence of polarization angle has been observed, revealing the crystal structure of bare 2L CrI$_3$ is still persisted in the monoclinic phase till to 90 K. Then, the temperature is further cooled down to 10 K to achieve magnetization. The fourfold dependence pattern is still striking (Figure 2(d)), validating that the crystal structure of bare 2L CrI$_3$ is always maintained in the monoclinic phase and no crystal phase transition takes place. Therefore, the intrinsic structure of bare 2L CrI$_3$ is monoclinic, which is independent of temperature and magnetization state. Furthermore, we plot the XX and XY polarization signals at selected 10, 90 and 295 K shown in Figure 2(e). Upon cooling process, both the two split $A_g$ and $B_g$ modes exhibit a significant blue-shift and the energy difference between them decreases from 1.83 to 1.52 cm$^{-1}$. In short, the bare 2L CrI$_3$ exhibits a monoclinic stacking structure with an AFM magnetic state.

## 2.3 Intrinsic monoclinic phase on 3-5L CrI$_3$

Next, we focus on studying the intrinsic structure of 3-5L CrI$_3$ and the effect of hBN-encapsulation on the crystal phase transition. Figure 3(a) shows the out-of-plane magnetization as a function of external magnetic field for representative 3-5L CrI$_3$. The layer-by-layer switching behavior is observed as increasing magnetic fields and the layer-numbers are determined by the number of plateaus under an external magnetic field of 2 T. The sharp plateau transition behavior observed in the magnetization curve indicates that there is no in-plane spin component. Due to interlayer AFM coupling, a magnetic hysteresis loop of the CrI$_3$ with odd-layer (3, 5L) can be observed near 0 T, while the net magnetization of even-layered (4L) CrI$_3$ is zero. We also firstly confirm the high-temperature crystal structure of bare and hBN-encapsulated 3-5L CrI$_3$ by polarized Raman spectroscopy. For bare 3-5L CrI$_3$, a fourfold dependence on polarization angle is manifested at 295 K (Figure S4(a)-(c)), which indicates that the intrinsic structure of bare 3-5L CrI$_3$ is monoclinic at room temperature. Similarly, we have systematically investigated the hBN-encapsulated 3-5L CrI$_3$, and experimental results show that the high-temperature phase of encapsulated few-layer samples are also monoclinic (Figure S4(d)-(f)). The bare and hBN-encapsulated 5L CrI$_3$ as a prototypical example intuitively show the optical contrast (Figure S5). The CrI$_3$ is completely sandwiched by the top and bottom hBN layers. The hBN-encapsulation layers can prevent air hydrolysis and maintain the stability of the sample. However, whether the hBN-encapsulation affects the structural phase transition remains elusive. As shown in Figure 3(b)-(d), the nearly same fourfold dependence on polarization angle at 10 K indicates that hBN-encapsulated 3-5L CrI$_3$ are monoclinic phases and no crystal phase transition occurs with temperature. Therefore, figuring out the low-temperature crystal structure of bare 3-5L CrI$_3$ is a key step for verifying the effect of hBN encapsulation layers on phase transition. Strikingly, the polarization-dependence Raman spectra of bare 3-5L CrI$_3$ at 10 K still depicts a fourfold dependence behavior, and the monoclinic phase is maintained during the whole temperature range, indicating the intrinsic lattice structure of bare few-layer CrI$_3$ is monoclinic stacking (Figure 3(e)-(g)). Temperature dependent XX and XY polarization Raman spectra of bare 5L CrI$_3$ have been carried out to understand the lattice variation tendency upon cooling over a wide range from 295 to 10 K shown in Figure S6(a). As the temperature decreases, the two split $A_g$ and $B_g$ modes simultaneously exhibit a blue-shift tendency and gradually converge in a linear manner. Besides, the sudden increase in linewidth of $A_g$ and $B_g$ modes indicates the occurrence of spin-lattice coupling,[33, 35, 43] and the transition temperature of ~50 K corresponds to the Curie temperature of 5L CrI$_3$ (Figure S6(b)-(c)). However, the energy difference of the two split modes of few-layer CrI$_3$ is still distinguished till to 10 K and they have not been merged, indicating monoclinic stacking is persisted with decreasing temperature at all time, rather than translating to rhombohedral stacking. Therefore, our experimental results validate that the intrinsic crystal structure of few-layer CrI$_3$ is monoclinic phase, whatever bare or hBN-encapsulated they are. However, as shown in Figure 3(h), it is worth noting that the extracted slopes of hBN-encapsulated CrI$_3$ are all smaller than that of corresponding bare CrI$_3$ with the same layer-number, demonstrating that hBN layers affect the lattice phase transition process.

The encapsulation hBN layers mitigate interlayer lateral sliding with temperature and partly persist the monoclinic stacking, but do not determine the intrinsic crystal structure of few-layer CrI$_3$. For sake of further studying the effect of hBN-encapsulation and guaranteeing that the hBN-encapsulation is the only variable, a half-encapsulated and half-bare 3L CrI$_3$ has been prepared, as shown in the inset of Figure 1(c), which ensures that the initial states from the bare and encapsulated parts are the same. As depicted in Figure S2, when the temperature drops from 295 to 10 K, for hBN-encapsulated area, the energy difference between $A_g$ and $B_g$ modes changes from 0.68 to 0.55 cm$^{-1}$. In stark contrast, the energy difference dramatically decreases from 0.65 to 0.4 cm$^{-1}$ for bare area. The change ratios of energy difference are 19% and 38% for the hBN-encapsulated and bare CrI$_3$, respectively. Thus, the change ratio of the bare area is twice as large as that of the hBN-encapsulated area, proving that hBN plays a spring damping effect, which hinders the interlayer sliding and prefers to persist monoclinic phase. In addition, the same comparison was made for a fully bare and a fully hBN-encapsulated 5L CrI$_3$ samples. Although the initial Raman mode energy differences are different due to the difference of the samples, similar results are also obtained, which proves the spring damping effect of hBN layers on lattice phase transition once again (Figure S7). Moreover, the slope difference between the bare and hBN-encapsulated samples shows a layer-dependence behavior, and the slope difference decreases with the increase of the number of sample layers, so it can be inferred that the spring damping effect of hBN layers mainly affects the surface layer of the sample, and therefore, the crystal structure of bulk sample is hardly affected by hBN layers because of its small surface-to-volume ratio[31, 44] (Figure 3(h)). In general, monoclinic stacking is the intrinsic structure of few-layer $CrI_3$, which has been always maintained with decreasing temperature and no lattice phase transition occurs; the hBN-encapsulation layers can partly diminish interlayer lateral sliding and benefit to keep monoclinic stacking.

## 2.4 Magnetic-field independence of crystal structure

A magnetic-field induced first-order structural phase transition from rhombohedral to monoclinic phase has been recently reported in bulk $CrI_3$ under a magnetic field of 2 T;[44] in stark contrast, an opposite interlayer sliding from monoclinic to rhombohedral stacking under magnetic fields has been observed in hBN-encapsulated bilayer $CrI_3$.[33] We further study the response of the lattice structure to the external magnetic field. A bare 3L $CrI_3$ flake has been chosen as a typical example. Temperature-dependent RMCD measurements reveal that the magnetization state of bare 3L $CrI_3$ remains up to ~50 K (Figure 4(a)). In Figure 4(b)-(c), in the case of non-magnetized state (90 K), the monoclinic stacking is maintained under both magnetic fields of 0 and 2.5 T; and thus, magnetic field has no effect on the crystal structure of 3L $CrI_3$. Subsequently, cooling down to 10 K and studying the effect of magnetic order; when applying a magnetic field of 2.5 T, the interlayer coupling changes from AFM to FM, and there is also an apparent fourfold dependence behavior in the polarized Raman spectrum, consistent with that at 0 T, presented in Figure 4(d)-(e). Raman spectra of XX and XY channels of bare 3L $CrI_3$ under several selected magnetic fields shown in Figure S8. As the magnetic field increases from 0 to +2.5 T (-2.5 T), the two split $A_g$ and $B_g$ modes remain nearly unchanged under both non-magnetized (90 K) and magnetized (10 K) states, indicating that there is no interlayer sliding and phase transition from monoclinic to rhombohedral stacking. Alternatively, 4L CrI$_3$ exhibits similar experimental results of magnetic-field-independence of crystal structure (Figure S9 and S10). These results validate that the external magnetic field and magnetic order hardly result in lattice phase transition. External magnetic field does not provide enough energy to overcome the structural energy barrier to achieve phase transition, although the energy barrier between the two stacking orders is only tens of mill-electron volts.[18, 33] On the contrary, a phase transition of structural stacking order can lead to a change of magnetic phase, indicating a unidirectional control behavior.

## 3 Conclusions

In summary, we have systematically investigated the intrinsic structure of few-layer bare CrI$_3$ and the effect of hBN on lattice phase transition. The hBN encapsulation layers are experimentally revealed to induce a spring damping effect, which break out the traditional wisdom that hBN is only used as a protection layer of 2D materials. But the intrinsic lattice phase of bare few-layer CrI$_3$ is essentially monoclinic, rather than hBN encapsulation layers, which just partly diminish interlayer sliding and facilitate to hold few-layer CrI$_3$ in monoclinic stacking. This work overturns the hypothesis that the monoclinic lattice phase of bare few-layer CrI$_3$ arise from hBN-encapsulation. Understanding the intrinsic lattice structure and magnetism and their correlation provide vital foundation for building novel 2D spintronic and memory devices.[45]

## 4 Methods

### 4.1 Sample Preparation

The few-layer CrI$_3$ were exfoliated from bulk crystal synthesized by chemical vapor transport method. We exfoliated bare few-layer CrI$_3$ onto SiO$_2$/Si substrates through PDMS films and also fabricated hBN/CrI$_3$/hBN heterostructure through 2D transfer technology, which were in-situ loaded into a closed cycle optical cryostat in glovebox.

### 4.2 RMCD and Raman Measurement

A 633 nm HeNe laser was coupled to Witec Raman system with closed cycle superconducting magnet and He optical cryostat. The light of ~6 μW was modulated by photoelastic modulator (PEM, $f_{PEM}$ = 50 KHz) and focused onto samples by a long working distance 50× objective (NA = 0.45). The reflected beams were collected by the same objective, passed through a non-polarizing beamsplitter cube into a photodetector. The Raman signals were recorded upon 532 nm light excitation by a Witec Alpha 300R Plus confocal Raman microscope. The polarization-resolved Raman spectra were obtained by rotating the half-wave plate in crossed-polarization configuration. The power of the 532 nm excitation laser was ~2 mW.

**Author contribution:** B.P., L.J.D. developed the concept, designed the experiment and prepared the manuscript. Z.L., Z.Y.C., Y.Z.G, T.G. Y.C.C. prepared the CrI3 samples. Z.L. performed the RMCD and Raman measurements. Z.L., H.P.L., Y.L., Y.T.N., B.P. contributed to the preparation of the manuscript.

**Research funding:** We acknowledge the financial support from National Science Foundation of China (51872039, 52021001, 51972046).

**Conflict of interest statement:** The authors declare no conflicts of interest regarding this article.

# Figures and captions

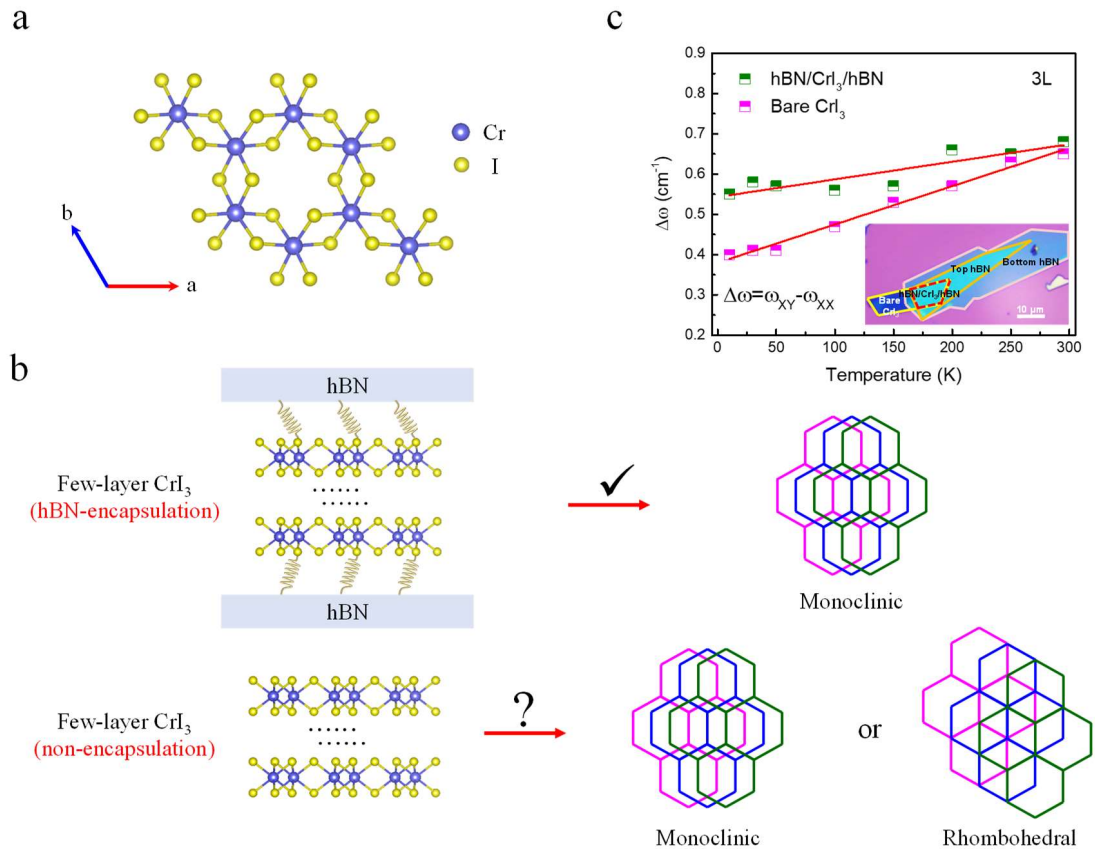

**Figure 1:** Schematic of intrinsic structure of few-layer CrI$_3$. (a) The in-plane atomic arrangement of monolayer CrI$_3$. The Cr atoms are coordinated to six I atoms and the adjacent Cr atoms form a honeycomb structure. (b) The schematic of crystal structure between hBN-encapsulated and bare few-layer CrI$_3$ ferromagnet. The hBN-encapsulated few-layer CrI$_3$ maintains monoclinic crystal structure through a spring damping effect, which has been hypothesized from the hBN-encapsulation layers, but so far, both this hypothesis and the crystal structure of bare few-layer CrI$_3$ still remains elusive. (c) The Raman mode energy difference as a function of temperature in a half-bare and half-hBN-encapsulated 3L CrI$_3$. Inset is the optical image of a half-encapsulated and half-bare 3L CrI$_3$.

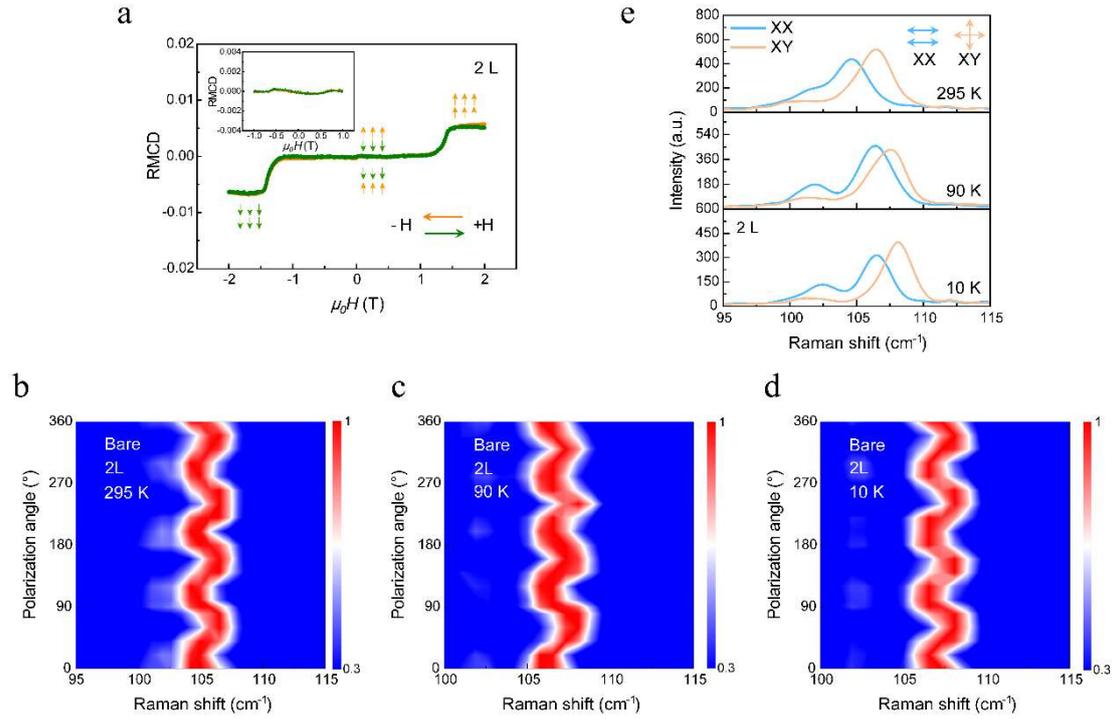

**Figure 2:** Intrinsic monoclinic phase of bare bilayer $CrI_3$. (a) RMCD signal of bilayer (2L) $CrI_3$ as a function of magnetic field. The orange (green) curve represents the magnetic field is swept from +2 (-2) to -2 (+2) T. Inset shows the RMCD spectrum of 2L $CrI_3$ within a magnetic field scope of ±1 T, no hysteresis loop and spin-flip state can be observed. (b)-(d) Polarized Raman spectra of bare 2L $CrI_3$ collected at 295, 90 and 10 K in the crossed-polarization configuration. A fourfold dependence of polarization angle can be observed, which evidently validates that the intrinsic lattice structure of bilayer $CrI_3$ is monoclinic throughout a whole temperature range. (e) Raman signals of 2L $CrI_3$ in the linearly parallel (XX) and crossed (XY) polarization channels taken at selected 10, 90 and 295 K.

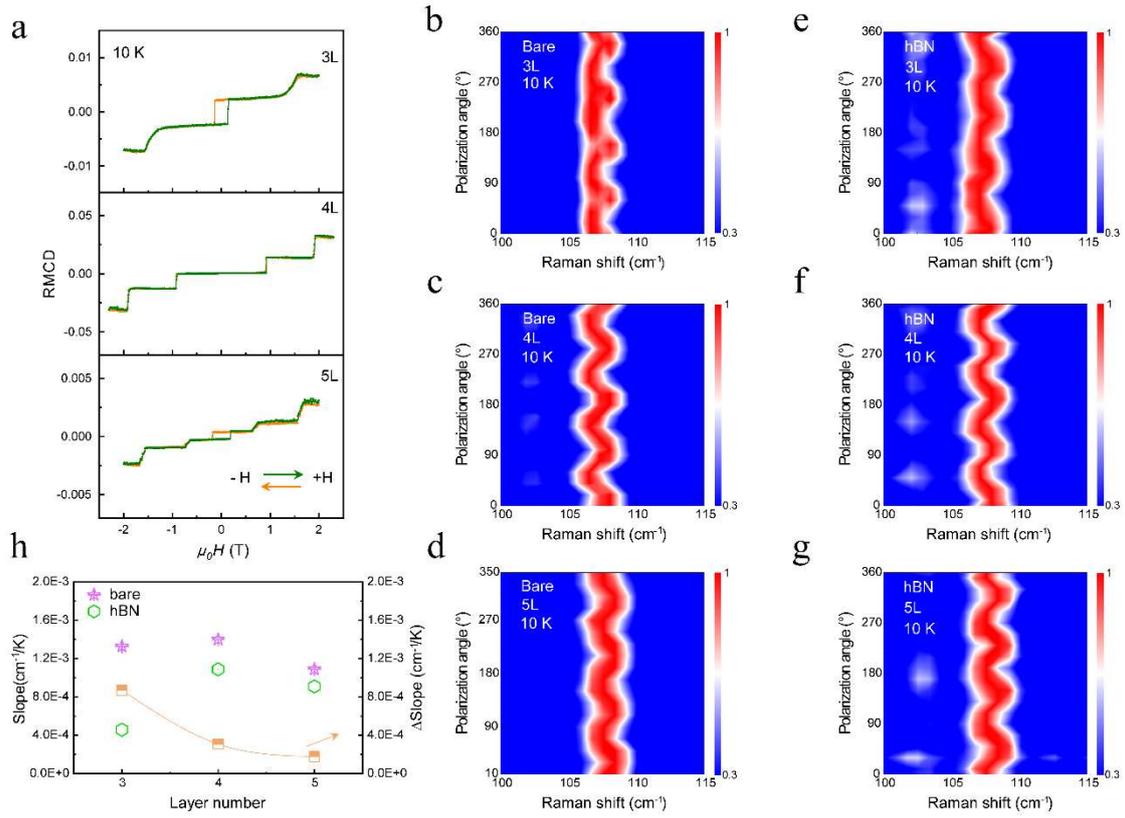

**Figure 3:** Intrinsic monoclinic phase of bare and encapsulated 3-5L $CrI_3$. (a) RMCD signals of 3-5L $CrI_3$ as a function of magnetic field taken at 10 K. The orange (green) curve represents the magnetic field is swept from +2 (-2) to -2 (+2) T. (b)-(d) Polarization angle dependence of Raman spectra of hBN-encapsulated 3-5L $CrI_3$ at 10 K. (e)-(g) Polarization angle dependence of Raman spectra of bare 3-5L $CrI_3$ at 10 K. Both of them show fourfold dependence patterns, in agreement with monoclinic stacking. (h) The extracted slopes of Raman mode energy difference of bare and hBN-encapsulated 3-5L $CrI_3$. The slopes of the hBN-encapsulated samples are lower than that of the corresponding bare samples with the same layer-number, indicating the spring damping effect of hBN layers on structural phase transition.

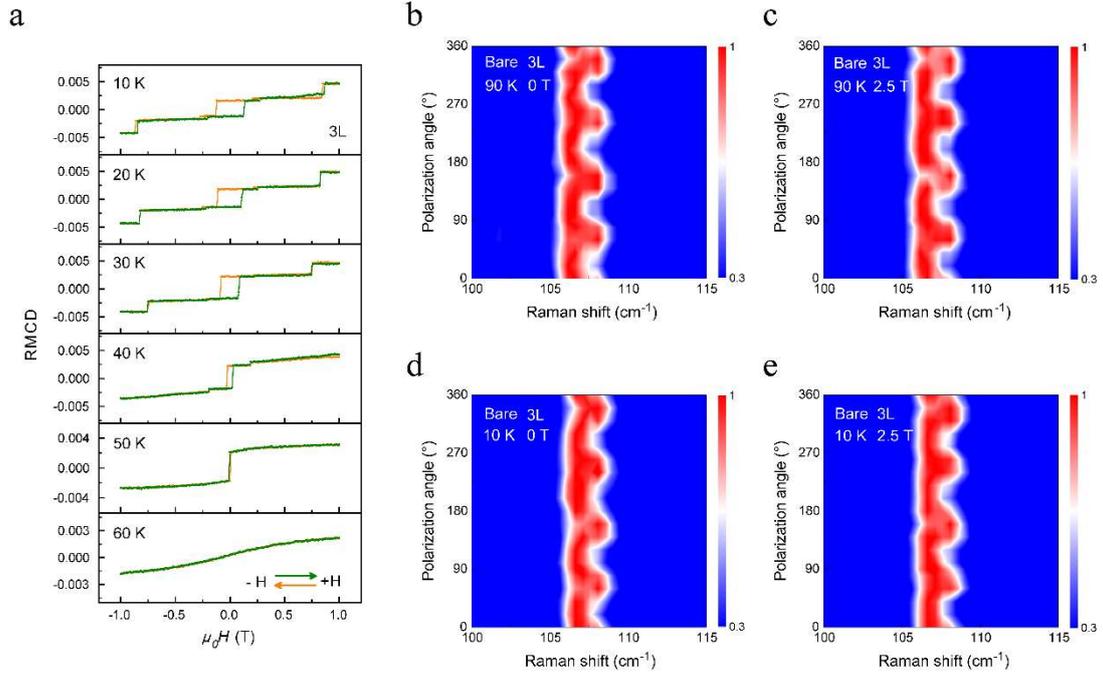

**Figure 4:** Magnetic-field independence of crystal structure. (a) Temperature-dependence RMCD spectra of bare 3L CrI$_3$. The magnetic hysteresis loop vanishes at 60 K, indicating a Curie temperature ($T_c$) is about 60 K. (b)-(c) Polarization angle dependence of Raman spectra of bare 3L CrI$_3$ collected at 0 and 2.5 T at 90 K. (d)-(e) Polarization angle dependence of Raman spectra of 3L CrI$_3$ collected at 0 and 2.5 T at 10 K.